\documentclass[useAMS,usenatbib]{mn2e}

\title[]{Curious Variables Experiment (CURVE):
SDSS J210014.12+004446.0 - dwarf nova with negative and positive superhumps}
\author[A. Olech et al.]{A. Olech$^{1}$\thanks{e-mail:
olech@camk.edu.pl},  A. Rutkowski$^{1}$ and
A. Schwarzenberg-Czerny$^{1,2}$\\
$^{1}$Nicolaus Copernicus Astronomical Center,
Polish Academy of Sciences, ul.~Bartycka~18, 00-716~Warszawa, Poland\\
$^{2}$A. Mickiewicz University Observatory, ul. S{\l}oneczna 36,
60-286 Pozna\'n, Poland}

\begin{document}

\date{Accepted 2008 December 15. Received 2008 November 30; in original form 2008 October 30}

\pagerange{\pageref{firstpage}--\pageref{lastpage}} \pubyear{2009}

\maketitle

\label{firstpage}

\begin{abstract}

We report the results of 67 hours of CCD photometry of the recently
discovered dwarf nova SDSS J210014.12+004446.0 (SDSS J2100). The data
were obtained on 24 nights spanning a month. During this time we
observed four ordinary outbursts  lasting about 2-3 days and reaching an
amplitude of $\sim$1.7 mag.  On all nights our light curve revealed
persistent modulation with the stable period of 0.081088(3) days
($116.767\pm0.004$ min) and  large amplitude of 0.5-0.6 mag in
quiescence reduced to 0.1-0.2 mag during outbursts.

These humps were already observed on one night by Tramposch et al.
(2005), who additionally  observed superhumps during a superoutburst.
Remarkably, from scant evidence at their disposal  they were able to
discern them as negative and positive (common) superhumps, respectively.
Our period in quiescence clearly different from their superhump period
confirmed this.  Our discovery of an additional modulation, attributed
by us to the orbital wave, completes the overall picture. Lack of
superhumps in our data indicates that all  eruptions we observed were
ordinary outbursts. The earlier observation of the superhumps combined
with the presence of the ordinary outbursts in our data enables
classification of SDSS J2100 as an active SU UMa dwarf nova with two
types of outbursts.

Additionally, we have promoted SDSS J2100 to the select group of
cataclysmic variables  exhibiting three periodic modulations of light
from their accretion discs. We updated available information on positive
and negative superhumps and thus provided enhanced evidence that their
properties are strongly correlated mutually as well as with the orbital
period. By recourse to these relations  we were able to remove an alias
ambiguity and to identify the orbital period of SDSS J2100  of
0.083304(6) days ($119.958 \pm 0.009$ min). SDSS J21000 is only third SU
UMa dwarf nova showing both  positive and negative superhumps. Their
respective period excess and deficit equal to $4.99\pm0.03$\% and
$-2.660\pm0.008$\%, yielding the mass ratio $q\approx 0.24$.

\end{abstract}


\section{Introduction}

In the zoo of variable stars SU UMa stars arguably belong to the most
intriguing exhibits. They are cataclysmic binaries, consisting of a
non-magnetic white dwarf and a late main sequence secondary filling its
Roche lobe and loosing mass through the inner Lagrangian point.  Their
orbital periods usually are less than the period gap at 2.5 hours. The
transferred matter forms an accretion disc around the primary (Warner
1995, Hellier 2001). In many cataclysmic variables accretion rate varies
considerably resulting in  brightness increase by several magnitudes
recurring in a semi-regular fashion after weeks or months intervals. The
bright phases are called {\em outbursts} and faint ones {\em
quiescence}.

The SU UMa stars are unique in that they simultaneously exhibit two
patterns of re-brightening: the {\em ordinary outbursts}, with
amplitudes typically 2-6 magnitudes, and {\em superoutbursts} brighter
by about one magnitude than the ordinary ones. In most cases the
superoutbursts recur more regularly at intervals several times longer
than the ordinary outbursts. All superoutbursts studied sufficiently
well reveal characteristic periodic tooth-shaped modulation called
superhumps. Their periods are couple hours and amplitudes are of order
0.1 magnitude. 

The peculiar behavior of SU UMa stars can be understood within the frame
of the thermal and tidal instability model (see Osaki 1996 for review).
The superhumps period is slightly longer than the orbital period of the
binary star. Most likely they are caused by prograde rotation of the
line of the apsides of a disk elongated by tidal perturbation from the
secondary. The perturbation is most effective when disk particles moving
in eccentric  orbits enter the 3:1 resonance with the binary orbit. Then
the superhump period is simply the beat period between orbital and
precession rate periods (Whitehurst 1988).

\begin{figure*}
\centering
\includegraphics{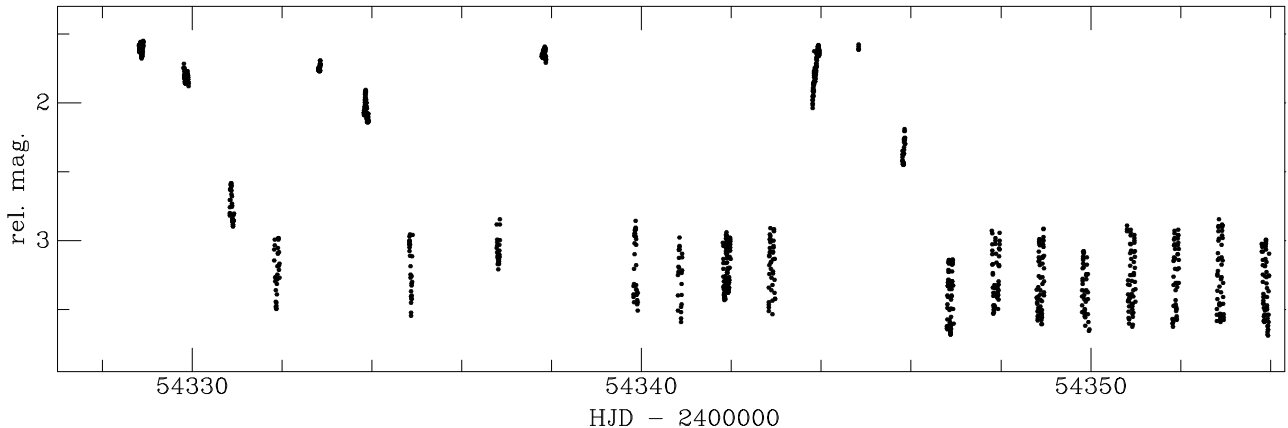}
\vspace*{6cm}
\caption{Global light curve of SDSS J2100 during our 2007 August-September campaign.}
\end{figure*}

Apart from the ordinary- and super-outbursts with the superhumps SU UMa
stars exhibit often additional types of periodic modulation. An orbital wave 
and/or eclipses may occur for inclination $i\ge60^\circ$. In the 
late stages of superoutbursts and during the early quiescence may appear
late superhumps, shifted by 0.5 in phase in respect to the
ordinary superhumps. In some systems negative superhumps occur with a period
slightly shorter than the orbital period. They are explained by invoking the 
classical retrograde precession of the disc tilted with respect to the
orbital plane (Wood \& Burke 2007).

On one hand the positive and negative superhumps were discovered together in just two
SU UMa stars, namely V503 Cyg (Harvey et al. 1995)
and BF Ara (Kato et al. 2003b, Olech et al. 2007). On the other hand,
such phenomena were already observed in AM CVn variables (AM CVn
itself - Skillman et al. 1999), classical novae (V1974 Cyg - Retter et
al. 1997 and V603 Aql - Patterson et al. 1997), VY Scl stars (TT Ari -
Skillman et al. 1998), SW Sex stars (V795 Her and DW UMa - Patterson et
al. 2002), nova-like variables (TV Col - Retter et al. 2003) and even
in the low mass X-ray binaries (V1405 Aql - Retter et al. 2002)

\section{SDSS J210014.12+004446.0}

After examining objects from Sloan Digital Sky Survey (SDSS) Szkody et
al. (2004) listed SDSS J2100 as a candidate dwarf nova because of its
spectrum characteristic for dwarf novae in outburst. The follow-up
photometric and spectroscopic observations of SDSS J2100 were obtained
by Tramposch et al. (2005). Their photometry was obtained on 3 nights
spanning 3 months and on each occasion extended over 3 cycles.  On two
nights the star was bright and clearly revealed the tooth-shaped
modulation with period of 2.099(2) hours (0.08746 days) and amplitude of
0.3 mag. On third night it was in quiescence and pulsating sinusoidally
with period 1.96(2) hours (0.0817 days) and amplitude reaching 0.5 mag.
Tramposch et al. (2005) argued that because of overall similarity of
SDSS J2100 to V503 Cyg (Harvey et al. 1995) the modulation in quiescence
 corresponds to the negative superhumps. However, light curves of
cataclysmic variables are known to suffer from red noise due to
flickering. Because of that any derived periods  have errors
underestimated, with no regard wheather they come from the least squares
or the white noise simulations (Schwarzenberg-Czerny, 1991).  This
combined with scant evidence at hand prompted us to look closer at this
star.

\section{Observations and data reduction}

Curious Variables Experiment (CURVE) is a long-term project of 
photometric observations of interesting variable stars in the Galaxy
and its clusters (Olech et al. 2008, Pietrukowicz et al. 2008). In order to study
southern objects we applied for time on the 1m telescope of the South African
Astronomical Observatory (SAAO). The time allocated to our project
lasted from August 15 till September 11, 2007.

\begin{table}
\centering
\caption{Journal of SAAO observations of SDSS J210014.12+004446.0}
\begin{tabular}{|l|c|c|c|r|}
\hline
\hline
Date of & HJD-2454000 & HJD-2454000 & Length & No. \\
2007    & Start       & End         & [h]    & exp. \\
\hline
\hline
Aug 16 & 329.31619 & 329.41052 & 2.264 & 76\\
Aug 17 & 330.30653 & 330.41395 & 2.578 & 79\\
Aug 18 & 331.33241 & 331.42551 & 2.234 & 41\\
Aug 19 & 332.32616 & 332.44182 & 2.776 & 31\\
Aug 20 & 333.30906 & 333.34797 & 0.934 & 75\\
Aug 21 & 334.32155 & 334.42363 & 2.450 & 89\\
Aug 22 & 335.33009 & 335.41185 & 1.962 & 29\\
Aug 24 & 337.27465 & 337.34748 & 1.748 & 30\\
Aug 25 & 338.28106 & 338.37153 & 2.171 & 59\\
Aug 27 & 340.32293 & 340.41477 & 2.204 & 34\\
Aug 28 & 341.30609 & 341.40181 & 2.297 & 24\\
Aug 29 & 342.31350 & 342.47927 & 3.978 & 99\\
Aug 30 & 343.32932 & 343.47112 & 3.403 & 45\\
Aug 31 & 344.30701 & 344.45856 & 3.637 & 121\\
Sep 01 & 345.32769 & 345.33317 & 0.132 & 5\\
Sep 02 & 346.30706 & 346.36444 & 1.377 & 35\\
Sep 03 & 347.30015 & 347.44513 & 3.480 & 67\\
Sep 04 & 348.30153 & 348.47059 & 4.057 & 52\\
Sep 05 & 349.29878 & 349.46284 & 3.937 & 64\\
Sep 06 & 350.30580 & 350.46436 & 3.805 & 41\\
Sep 07 & 351.30375 & 351.49162 & 4.509 & 58\\
Sep 08 & 352.31904 & 352.46207 & 3.433 & 47\\
Sep 09 & 353.30317 & 353.46041 & 3.774 & 49\\
Sep 10 & 354.30464 & 354.46907 & 3.946 & 64\\
\hline
Total   &     -     &    -      & 67.086 & 1314 \\
\hline
\hline
\end{tabular}
\end{table}

The telescope was equipped  with the STE3  camera with SITe CCD chip of
size $512\times 512$ pixels, back illuminated, cooled with liquid
nitrogen. At the focal length of 8.5m the scale was 0.31 arcsec/pix
providing the field of view of  $158 \times 158$ arcsec. The camera
design aims to minimize the noise from the bias and dark current hence
we skipped obtaining bias and dark frames. Although Johnson-Cousins
$UBV(RI)$ filters were fitted into filter wheel, most of our
observations were obtained through clear opening, in white light. This
enabled us to obtain good quality  photometry of faint objects and to
study their short time variations.  The exposure times ranged from 100
to 200 sec, depending on weather and actual brightness of the star. Our
data were reduced in a standard  way using procedures from the IRAF
package.\footnote{ IRAF is distributed by the National Optical Astronomy
Observatory, which is operated by the Association of Universities for
Research in Astronomy, Inc., under cooperative agreement with the
National Science Foundation.} The profile photometry has been derived
using the DAOphotII package (Stetson 1987). The differential photometry
yielded  typical accuracy of 0.01 mag or less.

Table 1 presents the journal of our observations of SDSS J2100. In
total, we obtained 1314 exposures in 67.1 hours of observations during 24 nights.

\begin{figure}
\centering
\includegraphics{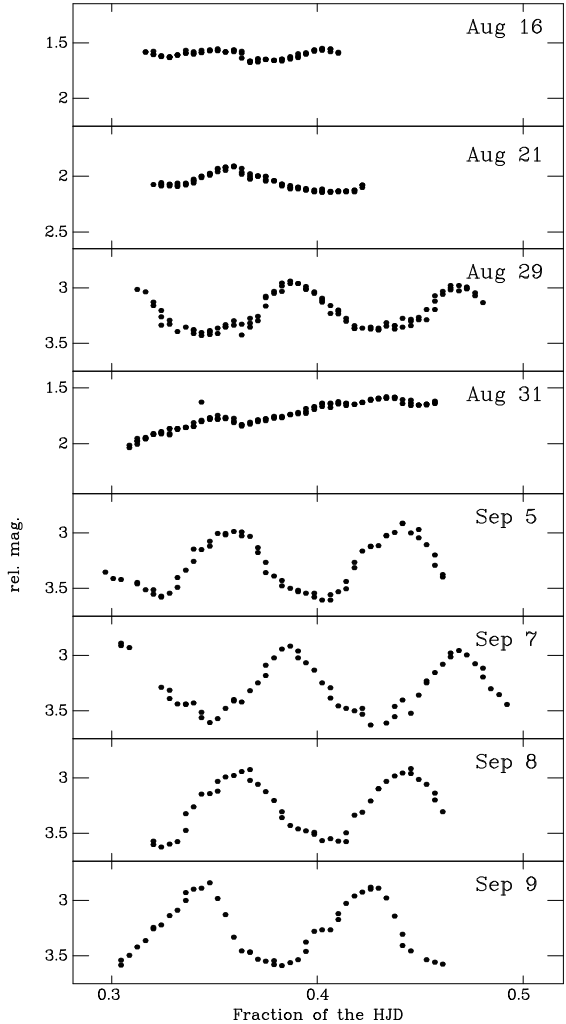}
\vspace{14cm}
\caption{Sample light curves of SDSS J2100 from outbursts and quiescence.}
\end{figure}

\section{Period analysis}

The global light curve, spanning almost one month of systematic
monitoring, is shown in Fig. 1. It demonstates that SDSS J2100 is an
active dwarf nova with frequent ordinary outbursts. We observed four
outbursts with a typical amplitude of $\sim$1.7 mag and duration of 2-3
days. In Fig. 2 we plot expanded sample light curves from eight
nights.  They  demonstrate behaviour of the star both in outbursts (on
Aug 16, Aug 21 and Aug 31) and during quiescence (Aug 29, Sep 5-9).
Inspection of the light curve reveals clearly presence of the modulation
with a period of around two hours and with the amplitude changing from
0.1-0.2 mag in outburst to 0.5-0.6 mag during quiescence.

To facilitate our period analysis, we transformed the light curve of
SDSS J2100 in two ways.  First, we converted it from the magnitudes into
corresponding intensity units.  In this way the amplitudes of modulation
on all nights became comparable.  Next, each night light curve was
detrended by subtraction of the best fit parabola. In this way we
obtained a light curve expressed in counts with its mean intensity set
at zero. The ANOVA periodogram (Schwarzenberg-Czerny 1996) computed  for
the result light curve is shown in Fig. 3 and refered as $A$.

\begin{figure}
\centering
\includegraphics{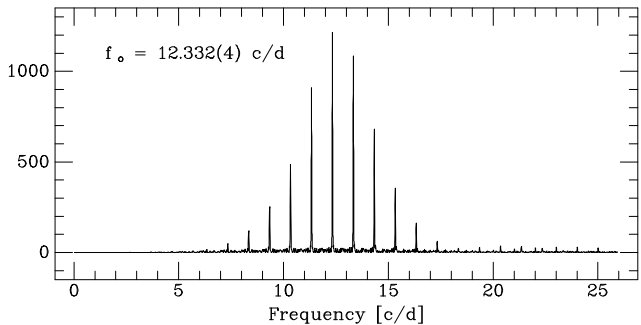}
\vspace{4.5cm}
\caption{ANOVA periodogram for detrended intensity light curve of SDSS J2100.}
\end{figure}

The periodogram $A$ is dominated by a window pattern with its peak centered
at frequency $f_0 = 12.3321(40)$ c/d accompanied with 1 c/d aliases. The
peak value corresponds to the period of 0.08109(3) days and agrees with
the period of 0.0817(8) days reported by Tramposch et al. (2005) during
quiescence.

The fact that the 0.0811-day modulation revealed changes both in shape
and amplitude raised suspicion of presence in our light curve of yet
another periodicity. To check this hypothesis we prewhitened our data
with the main periodicity and its two harmonics. The resulting periodogram $B$
looks rather complex but clearly reveals  presence of some remaining oscillation.

\begin{figure}
\centering
\includegraphics{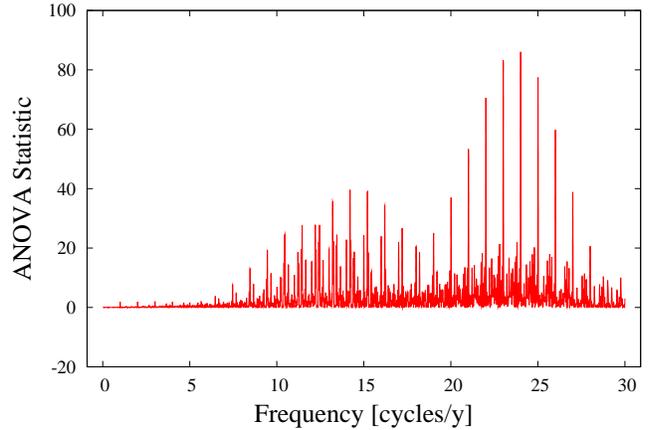}
\vspace{6cm}
\caption{ANOVA periodogram for light curve of SDSS J2100 prewhitened with
0.0811-day oscillation of variable amplitude (see text).}
\end{figure}

In the periodogram $B$ there are at least two overlapping window patterns.
At high frequencies peaks cluster close to- but not quite at- the $2 f_0$ harmonic. 
Still, much power is left at low frequencies. A pronounced pattern centered near the
removed frequency $f_0$ remains. It 
contains a stong side-band doublet centered exactly at $f_0$. This is consistent with remains  of
original peak at $f_0$ broadened due to modulation of amplitude and/or phase and demonstrates
insufficient prewhitening in the periodogram $B$. 

We turn back to the original data to better remove any interference from the $f_0$ modulation. In our second try we prewhiten the original data
with a sine function of fixed frequency $f_0$ {\em and} of a modulated amplitude.
See Appendix \ref{a2} for the description of our procedure.
The periodogram $C$ resulting from our second attempt of prewhitening of $f_0$ is shown in Fig. 4. 
Now the pattern at low frequencies is markedly fainter and the periodogram is dominated at high frequency by a pattern centered around its peak at $f_1=24.008(2)$ c/s. 
The $\pm 1$ cycle/day aliases at $f_2=23.010(2)$ and $f_3=25.007(2)$ c/d are prominent and we could not exclude at this stage that one represents the true frequency. The detailed values were obtained by least squares fit. The corresponding periods are $P_1=0.041652(3)$ days, $P_2=0.043460(3)$ days and $P_3=0.039989(3)$ days, respectively.
In the periodogram $B$ the dominant peak corresponded to $f_2$. One could argue that in that case because of the presence of the strong low frequency pattern, the relative height of the high frequency peaks was distorted due to interference with the residues of the $2 f_0$ harmonic. This argument is further supported by consideration of errors of sine fits of the new frequencies. For the case $B$ errors of $f_2$ and $f_1$ were factor 4 larger than those quoted above for the case $C$. 
Prewhitening of $f_1$ leaves no evidence of any remaining periodic modulation. 

It may be informative to present our conclusion in purely statistical terms. For our periodogram $C$ the peak values of the ANOVA statistics exceed 80 for 3 and 1314 degrees of freedom. Accordingly, the null hypothesis stating that our data reveals no coherent periodic modulation must be rejected at the confidence level of $>10\sigma$. In that sense our detection of the secondary periodicity in SDSS J2100 is secure. The remaining alias ambiguity concerns of the frequency value and not of the reality of the observed modulation.

\begin{figure}
\centering
\includegraphics{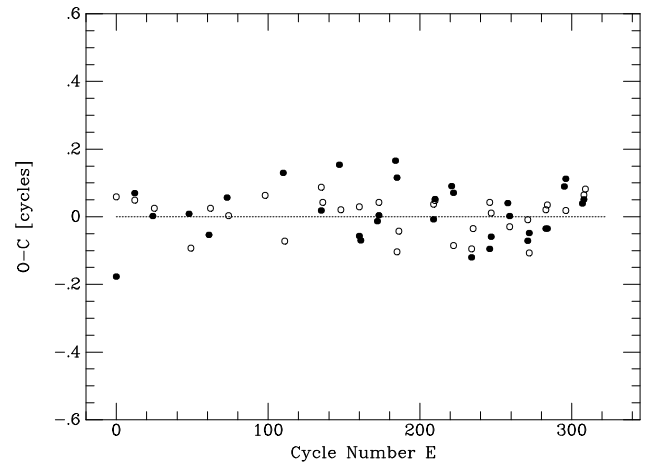}
\vspace{6.4cm}
\caption{$O-C$ values for 0.0811-day periodicity of SDSS J2100. Dots and
open circles corresponds to the maxima and minima, respectively.}
\end{figure}

\section{O-C analysis}

The light curve of SDSS J2100 from August and September 2007 contains 32
maxima and 30 minima. To check for any modulation of the light curve shape 
we examined them separately. The cycle numbers $E$, times and errors and  $O-C$ values 
are listed in Table 2. The residuals $O-C$ are computed using 
the ephemerides given below.

\begin{table}
\centering
\caption{Times of maxima and minima in the light curve of SDSS J2100.}
\begin{tabular}{|r|c|r||r|c|r|}
\hline
\multicolumn{3}{|c|}{Maxima} & \multicolumn{3}{|c|}{Minima} \\
\hline
$E$ & ${\rm HJD}_{\rm max}$ & $O-C$     & $E$ & ${\rm HJD}_{\rm min}$ & $O-C$ \\
\hline
  0 & 29.4040(50) & $-0.176$  &   0 & 29.3750(40) &  $0.059$\\
 12 & 30.3970(35) &  $0.070$  &  12 & 30.3475(60) &  $0.049$\\
 24 & 31.3645(40) &  $0.003$  &  25 & 31.4000(45) &  $0.026$\\
 48 & 33.3110(50) &  $0.009$  &  49 & 33.3370(35) & $-0.093$\\
 61 & 34.3600(25) & $-0.054$  &  62 & 34.4010(50) &  $0.025$\\
 73 & 35.3420(40) &  $0.057$  &  74 & 35.3725(35) &  $0.003$\\
110 & 38.3480(45) &  $0.131$  &  98 & 37.3240(60) &  $0.063$\\
135 & 40.3660(45) &  $0.018$  & 111 & 38.3675(40) & $-0.071$\\
147 & 41.3500(80) &  $0.154$  & 135 & 40.3270(35) &  $0.087$\\
160 & 42.3870(30) & $-0.056$  & 136 & 40.4045(70) &  $0.043$\\
161 & 42.4670(40) & $-0.070$  & 148 & 41.3760(55) &  $0.021$\\
172 & 43.3635(30) & $-0.013$  & 160 & 42.3500(60) &  $0.029$\\
173 & 43.4460(40) &  $0.004$  & 173 & 43.4055(45) &  $0.042$\\
184 & 44.3510(40) &  $0.166$  & 185 & 44.3670(50) & $-0.103$\\
185 & 44.4280(50) &  $0.115$  & 186 & 44.4530(30) & $-0.043$\\
209 & 46.3640(50) & $-0.008$  & 209 & 46.3250(50) &  $0.037$\\
210 & 46.4500(40) &  $0.053$  & 210 & 46.4070(50) &  $0.048$\\
221 & 47.3450(35) &  $0.091$  & 222 & 47.3695(50) & $-0.085$\\
222 & 47.4245(30) &  $0.071$  & 234 & 48.3420(45) & $-0.095$\\
234 & 48.3820(45) & $-0.120$  & 235 & 48.4280(70) & $-0.035$\\
246 & 49.3570(40) & $-0.095$  & 246 & 49.3265(30) &  $0.043$\\
247 & 49.4410(35) & $-0.059$  & 247 & 49.4050(35) &  $0.011$\\
258 & 50.3410(35) &  $0.041$  & 259 & 50.3750(45) & $-0.030$\\
259 & 50.4190(55) &  $0.003$  & 271 & 51.3500(30) & $-0.009$\\
271 & 51.3860(20) & $-0.071$  & 272 & 51.4232(30) & $-0.107$\\
272 & 51.4690(20) & $-0.048$  & 283 & 52.3257(40) &  $0.021$\\
283 & 52.3620(25) & $-0.034$  & 284 & 52.4080(40) &  $0.035$\\
284 & 52.4430(25) & $-0.035$  & 296 & 53.3800(40) &  $0.019$\\
295 & 53.3450(30) &  $0.089$  & 308 & 54.3570(40) &  $0.064$\\
296 & 53.4280(30) &  $0.113$  & 309 & 54.4395(35) &  $0.082$\\
307 & 54.3140(50) &  $0.040$  &  &  & \\
308 & 54.3960(40) &  $0.051$  &  &  & \\
\hline
\end{tabular}
\end{table}
\medskip

The moments of maxima and minima can be fitted with 
their respective linear ephemerides of the form:

\begin{eqnarray}
{\rm HJD}_{\rm max} &=& 2454329.4183(16) + 0.081083(7) \cdot E\\
{\rm HJD}_{\rm min} &=& 2454329.3702(17) + 0.081109(8) \cdot E
\end{eqnarray}

The $O-C$ values computed according to the above-mentioned formulae
are presented in Table 2 and also plotted in Fig. 5. They are consistent with
the 0.0811-day period remaining constant during our run.

To calculate the final value of the principal negative hump period of SDSS J2100 we fitted
the light curve with a sinusoid, yielding $P_0=0.081088(3)$ days ($116.767\pm0.004$ min).
This period differs only by 0.000006 days from the weighted mean of the values from 
Eqs. 1 and 2, suggesting that its error is not underestimated by 
more than a factor of 2 (Schwarzenberg-Czerny, 1991).

\section{Properties of SDSS J2100}

\subsection{The superhump period}

In classical SU UMa stars, the superhumps are observed during supermaxima,
their shape resembles shark teeth and their period is few percent longer than 
the orbital one while typical amplitudes are in the range 0.1-0.3 mag.
Curiously, this was exactly what Tramposch et al. (2005) have observed 
in SDSS J2100 during the July 2003 outburst. Namely, during
two nights the star was bright ($V\approx 16.3\div 16.5$ mag) and
revealed teeth-shaped oscillations with the period of $2.099\pm0.002$ 
hours i.e. 0.08746(8) days and the amplitude of 0.2-0.3 mag.
This and their lack during our outbursts confirms they were 
common (positive) superhumps arising from the apsidal precession 
of the eccentric accretion disc.

\subsection{The negative superhumps}

Some cataclysmic variables reveal the so-called negative superhumps. 
They are thought to occur due to the nodal precession of a tilted accretion disc. 
According to Wood \& Burke (2007) for the tilted accretion disk 
the stream often misses its rim, penetrating deeper inwards on most orbital phases. 
When the stream eventually collides with the disk, it releases more energy in the ensuing hot spot. 
Early collision occurs only when the stream sweeps across the disk rim (the nodal line). 
Depending whether the stream hits or misses the rim, as the hot spot migrates in and out,
and its brightness is modulated accordingly, roughly twice per each binary revolution. 
However, the observer sees only one event per cycle, from the hot spot occurring on 
the visible side of the disk. To complete the scenario, the tilted disc 
is subject to a slow retrograde precession, resulting in the modulation period
slightly shorter than the orbital one. 

The negative superhumps were detected so far just in two classical SU UMa stars
namely in V503 Cyg (Harvey et al. 1995) and BF Ara (Olech et al. 2007). In both stars 
the modulation persists over long period of time covering several cycles and 
supercycles and its amplitude in quiescence is large, reaching up to 0.5-1 mag.
Unlike in ordinary superhumps, their decline branches are often steeper than the rise.

Th 0.0811 day oscillation of SDSS J2100 described in Sect. 4 has all 
properties of the negative superhump. This is best appreciated by inspection of 
the light curve of SDSS J2100 phased and
binned with this period (Fig. 6). Each bin contains about 20 measurements and 
small scatter indicates excellent phasing. To make this picture clear,
we subtracted from the light curve the high frequency modulation, 
to be discussed in the next section.

\begin{figure}
\centering
\includegraphics{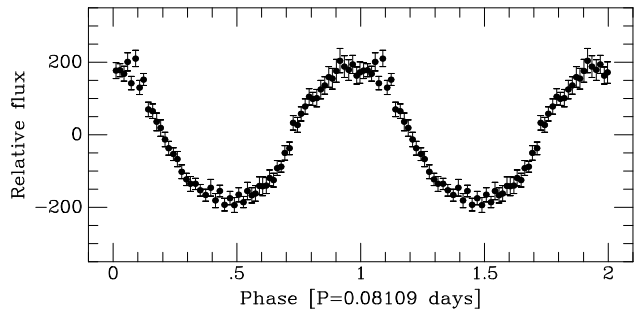}
\vspace{4.8cm}
\caption{Binned light curve of SDSS J2100 created after prewhitening with
high frequency signal $f_1$ and phased with period of 0.0811 days.}
\end{figure}

Note, that as pointed by Tramposch et al. (2005) if the 0.08746 day 
period corresponds to the positive superhump it would be wrong to attribute 
the 0.0811 day period to the orbital motion. The corresponding period 
excess $\epsilon$, defined as $P_{\rm sh}/P_{\rm orb}-1$,
of 8\% would exceed by a factor of two the values observed in other SU
UMa stars of the similar period. As discussed in Sect. 7, 
among SU UMa stars the period excess is strongly correlated with the orbital period.
In the next section we confirm this argument by detection of the separate orbital modulation.

\subsection{The orbital wave}

In quiescence dwarf novae of intermediate inclination often reveal
orbital modulation of their light curve. The modulation is caused by
changing aspect of the  non-transparent hot spot. Depending on the
transparency of the spot, the resulting light curve is complex, often
double humped. The double humped light curve reveals more power in its
$P_{\rm orb}/2$ harmonics than  in the fundamental period. In this
regard note that our light curve of SDSS J2100 after prewhitening  of
the negative superhump yields another modulation of frequency
$f_1=24.008$ c/d, about right  for the orbital harmonics. Thus $f_1/2=12.004$ c/d
would correspond to the orbital period of $P_{\rm
orb} = 0.083304(6)$ days. 
Recalling large power in 1 c/d alias $f_2$ 
it seems justified to consider it as an alternative orbital harmonics. The hitch is
that the orbital frequency $f_2/2 = 11.5049$ c/d corresponding to $P_2 =
0.08692$ days would yield a period excess of only 0.6\% i.e. much, much
smaller  than observed in the typical SU UMa stars of the orbital period
close to $P_2$. 

For comparison in Fig. 7 we plot the light curve of SDSS
J2100 prewhitened with the negative superhump modulation, phased and
binned with the periods of 0.08330 and 0.08692 days. Note that the
former period yields the largest peak-to-peak modulation, while the
latter one yields a smoothed-out light curve with similar maxima. 
This suggests that for the period 0.08692 alternate maxima of differing shapes
were averaged out and it lends further support to our orbital period of
0.08330 days. The asymmetry in the heights of
the maxima of the orbital light curve is consistent with the orbital wave hypothesis for a
semi-transparent hot spot. The higher maximum arises from the hot spot
visible face-on. The secondary maximum is produced half orbital cycle
later while only a fraction of radiation penetrates through the back of
the hot spot moved to the opposite side of the disc. 

\begin{figure}
\centering
\includegraphics{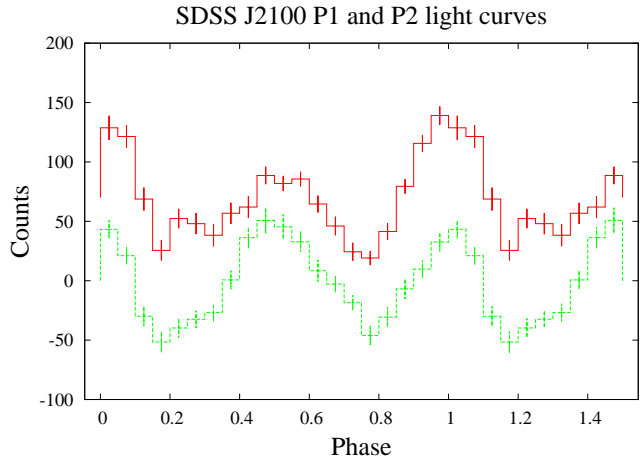}
\vspace{7cm}
\caption{The light curves of SDSS J2100 after prewhitening with
the negative superhumps signal $f_0$ phased with periods of 0.08330 and 0.08692 days,
top and bottom, respectively. To prevent overplotting we shifted them vertically.}
\end{figure}

\section{SDSS J2100 and its relatives}

\subsection{SDSS J2100 place in the bi-humpers family}

Summarizing conclusions from previous sections, SDSS J2100 reveals typical 
characteristic of an active SU UMa stars. Tramposch et al. (2005) likely
caught it during superoutburst and demonstrated presence of 
superhumps with the period of $P_{\rm sh}=0.08746(8)$ days ($125.94 \pm
0.12$ min). We observed frequent normal outbursts lasting 2-3 days and
with amplitudes of $\sim$1.7 mag.

Except for the superoutbursts, the light curve is dominated by the
combination of two oscillations with periods $P_{\rm ns}=0.081088(3)$
days ($116.767\pm0.004$ min) and $P_{\rm orb}=0.083304(6)$ days
($119.958\pm0.009$  min), interpreted as the negative superhumps and the
orbital period of the binary, respectively. Already Tramposch et al.
(2005) speculated that the three humps  seen by them on one night in
quiescence were negative superhumps. We refined their period and
discovered the orbital modulation, thus securing interpretation of all
three periods. Their presence allow us to determine  both the period
excess and the period deficit, respectively for the  positive and
negative humps, as equal to $\epsilon = 0.0499(32)$ and
$\epsilon_-=-0.02660(8)$.

The most recent compilation of the properties of stars showing
superhumps were made by Patterson (1998), Patterson et al. (2005) and
Pearson (2006). However, the last two papers contain only tables of values
without particular references for each object. This compelled us to redo
the literature survey again and to update it till the end of 2008.
The survey returned 112 cataclysmic
variables with known orbital and any superhump period (no matter
positive, negative or both). The complete material will
be discussed elsewhere (Olech 2009, in preparation). Its subset for the 
systems revealing simultaneously positive and negative superhumps is discussed in Sect. 7.2. 
Here in Fig. 8 we present only an updated graph of the period excesses/deficits, 
for all surveyed stars with $P_{\rm orb}>0.05$, plotted against their orbital period.

\begin{figure}
\centering
\includegraphics{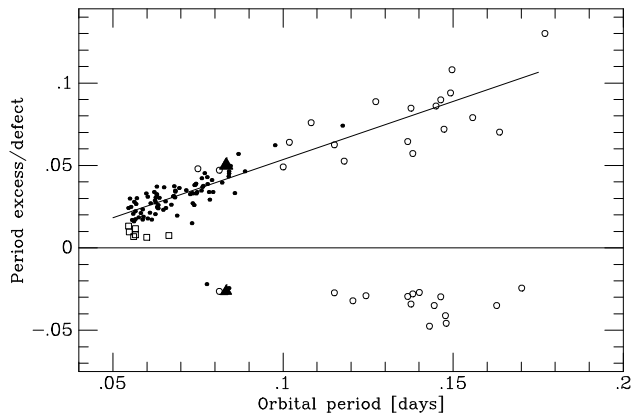}
\vspace{6cm}
\caption{Dependence between period excess or deficit and orbital period of
the binary for different types of cataclysmic variables. Ordinary SU UMa
stars are shown with dots. Open circles represent nova-like variables and
classical novae. The candidate period bouncers are plotted with open
squares. The position of SDSS J2100 corresponding to the adopted orbital period of
0.0833 day is marked with the solid triangles.}
\end{figure}

The big solid triangles in Fig. 8 correspond to the values of period
excess and deficit  for SDSS J2100, $\epsilon$ and $\epsilon_-$,
calculated for the orbital period  determined in Sect. 6.3. Their
excellent match with other points renders strong support to our choice
of the true orbital period among its aliases. For alias periods, the 
triangles would be located far away from the general trend. 

Pearson (2006) fitted a simple formula to the empirical relation between
period excess and mass ratio:

\begin{equation}
\epsilon = -4.1 \times 10^{-4} + 0.2076q
\end{equation}

\noindent By employing it we estimated the mass ratio for SDSS J2100 as equal
to $q=0.24$. Next, assuming that the secondary size is consistent
with the main sequence value we could employ the empirical mass-period
relation (Warner 1995):

\begin{equation}
M_2 = 0.065 P_{\rm orb}^{5/4}{\rm [h]}
\end{equation}
to derive masses. This yields tentative mass estimates $M_2=0.15$ and $M_1=0.64 M_{\odot}$.
This values are consistent with those for other dwarf novae just below of the 
orbital period gap between 2 and 3 hours.

\subsection{Properties of the bi-humpers} 

Retter et al. (2002) listed seven cataclysmic variables exhibiting
both positive and negative superhumps. Since then the number of such objects doubled, hence
we felt compelled to update their Table 2 with the object studied 
till 2008. These updated results are summarized in our Table 3.

\begin{table*}
\centering
\caption{Properties of the systems that have two kinds of superhumps. Stars
marked by asterisk have uncertain detections. For explanation of $S$,
$E$, $C$ and $L$ symbols see Appendix A.}\label{t3}
\begin{tabular}{|l|l|l|l|l|l|l|c|}
\hline
Object & Orbital & Positive & ~~~~~~$\epsilon$ & Negative & ~~~~~~$\epsilon_-$ & 
~$\phi=\epsilon_-/\epsilon$ & Ref. \\
       & Period [d] & Superhump [d] &             & Superhump [d] & & \\
\hline
AM CVn       & 0.011906623(3)$^E$ & 0.0121667(13)$^S$  & 0.02184(11)  & 0.01170613(35)$^E$ & $-0.01684(3)$ & $-0.771(4)$ & 1,2\\
V1405 Aql    & 0.034729754(11)$^E$ & 0.035041099(60)$^E$& 0.008965(18) & 0.034483(13)$^E$  & $-0.0071(4)$    & $-0.793(42)$ & 3,4\\
V1159 Ori$^*$& 0.06217801(13)$^C$ & 0.064167(40)$^E$  & 0.0320(6)   & 0.0583(3)$^S$   & $-0.0624(50)$   & $-1.95(15)$   & 5,6\\
ER UMa$^*$   & 0.06366(12)$^E$  & 0.065552(25)$^E$  & 0.030(2)    & 0.0589(7)$^E$     & $-0.075(11)$    & $-2.52(41)$   & 6,7,8\\
V503 Cyg     & 0.0777(8)$^E$    & 0.08104(7)$^E$  & 0.043(11)   & 0.07569(7)$^S$  & $-0.026(10)$     & $-0.60(28)$   & 9\\
V1974 Cyg    & 0.08125873(23)$^L$ & 0.0849(1)$^S$ & 0.0448(12)   & 0.07911(5)$^?$  & $-0.0264(6)$    & $-0.590(21)$  & 10,11,12,27\\ 
SDSS J2100   & 0.083304(6)$^L$    & 0.08746(27)$^E$  & 0.0499(32)  & 0.081088(3)$^L$  & $-0.02660(8)$   & $-0.533(35)$  & 13,14\\
BF Ara       & 0.084176(21)$^L$   & 0.08797(1)$^E$   & 0.0451(3)   & 0.0821(1)$^S$  & $-0.0247(12)$    & $-0.547(27)$  & 15,16\\
V592 Cas     & 0.115063(5)$^E$    & 0.12228(1)$^E$   & 0.06272(10)  & 0.11193(5)$^E$  & $-0.0272(4)$    & $-0.434(7)$   & 17\\
DW UMa       & 0.136606499(3)$^E$ & 0.14539(13)$^E$ & 0.0643(95)    & 0.13259(5)$^E$  & $-0.0294(4)$    & $-0.457(9)$   & 18,19\\
TT Ari       & 0.13755040(37)$^{C+S}$& 0.14926(5)$^L$  & 0.0851(4)   & 0.1329(3)$^S$    & $-0.0338(22)$     & $-0.397(26)$  & 20,21,28,29\\
V603 Aql     & 0.1381(1)$^E$   & 0.1460(7)$^S$   & 0.0572(51)   & 0.1343(3)$^S$   & $-0.0275(23)$     & $-0.481(59)$  &  22\\
RR Cha$^*$   & 0.1401(1)$^E$   & 0.1444(3)$^E$   & 0.0307(23)   & 0.1363(3)$^E$   & $-0.0271(23)$   & $-0.883(98)$  &  23\\
PX And       & 0.1463527(1)$^E$ & 0.1595(2)$^E$   & 0.0898(14)   & 0.141(1)$^S$    & $-0.0366(68)$   & $-0.407(76)$  & 24,25\\
TV Col       & 0.22860(1)$^S$   & 0.2639(35)$^L$  & 0.154(15)    & 0.2160(5)$^S$   & $-0.0551(22)$       & $-0.357(38)$  & 26\\
\hline
\multicolumn{8}{|l|}{Refs: 1 - Harvey et al. (1998), 2 - Skillman et al. (1999), 3 - Chou et al. (2001), 
4 - Retter et al. (2002), 5 - Patterson et al. (1995),}\\
\multicolumn{8}{|l|}{6 - Thorstensen et al. (1997), 7 - Kato et al. (2003a), 8 - Gao et al. (1999), 
9 - Harvey et al. (1995), 10 - Retter et al. (1997),}\\
\multicolumn{8}{|l|}{11 - Skillman et al. (1997), 12 - Olech (2002), 
13 - Tramposch et al. (2005), 14 - this work, 15 - Kato et al. (2003b),}\\
\multicolumn{8}{|l|}{16 - Olech et al. (2007), 17 - Taylor et al. (1998), 18 - Patterson et al. (2002), 
19 - Araujo-Betancor et al. (2003),}\\
\multicolumn{8}{|l|}{20 - Thorstensen et al. (1985), 21 - Skillman et al. (1998), 22 - Patterson et al. (1997), 
23 - Wouldt \& Warner (2002),}\\
\multicolumn{8}{|l|}{24 - Patterson (1998), 25 - Stanishev et al. (2002), 26 - Retter et al. (2003), 27 - Patterson (1999), 28 - Wu et al. (2002)}\\
\multicolumn{8}{|l|}{29 - Semeniuk et al. (1987)}\\
\end{tabular}
\end{table*}

Apart from the orbital and superhump periods we also list period
excesses and deficits and their ratio $\phi$. Retter et al. (2002) already suggested 
that $\phi$ correlates with the orbital period. Our Fig. 9, containing twice as much points
confirms their hypothesis. A straight line to fitted to the plotted the points yields the following
empirical formula:

\begin{figure}
\centering
\includegraphics{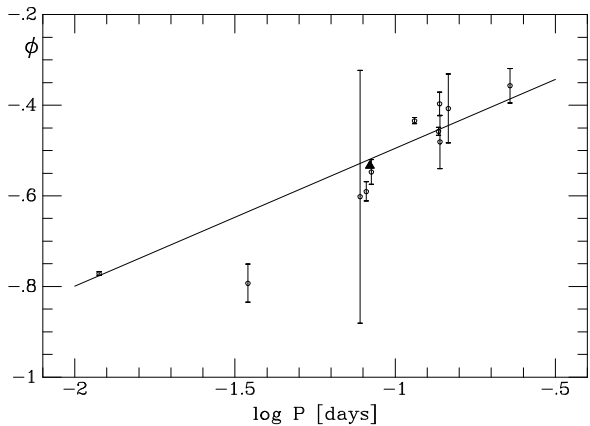}
\vspace{6cm}
\caption{Relation between the ratio between period deficit and excess and orbital period of
the binary for different types of cataclysmic variables. 
We did not plot V1159 Ori, ER UMa and RR Cha due to the uncertain detections of
the superhumps in these systems. Straight line
corresponds to the fit given by equation (4)}
\end{figure}

\begin{equation}
\phi = 0.318(6)\cdot\log P - 0.161(10)
\end{equation}

As was pointed out by Retter et al. (2002) existence of this relation poses some challenge to the
theory as it fits binaries with rather different components. It is obeyed by the double-degenerate
systems of AM CVn type, by the classical red dwarf - white dwarf cataclysmic binaries  and
by the low-mass X-ray binary containing a neutron star. This star-type independence seems to 
confirm beliefs that the positive and  negative superhumps are pure accretion disk phenomena, 
independent of any stellar influence except for the gravity.

\section*{Acknowledgments}

We acknowledge generous allocation of the SAAO 1-m telescope time. We
would like to thank Prof. J\'ozef Smak for fruitful discussions. This
work was supported by SALT Foundation and MNiSzW grant no. N N203 301335 to A.O.
and MNiSzW grant no. N~N203~3020~35 to A.S.-C.

\appendix
\section{Period errors}\label{a1}
It may be useful to summarise here principles of our review in Table \ref{t3} of literature on CVs' periods and their errors. The review is conservative in that whenever our error estimates differed from the original ones by less than factor 3 or our periods differed from the original ones by less than their error, we sticked with the respective numbers published by the original authors.
\subsection{Statistical errors}
\subsubsection{Rayleigh criterion}
In absence of any detailed information the statistical error of period of a CV may be estimated roughly as
\begin{equation}
\sigma_P = \frac{P^2\sigma_{\phi}}{T}\label{ea.1}
\end{equation} 
where $P$, $\sigma_{\phi}$ and $T$ denote respectively period, phase error and interval of observation.
For $\sigma_{\phi}=1$ Eq. (\ref{ea.1}) reduces to Rayleigh criterion.
 and yields maximum error consistent with the secure cycle count over the interval $T$. However, for the photometric observations of humps in CVs we adopt $\sigma_{\phi}=0.07$ to better reflect the phase indeterminacy due to flickering. For the orbital periods derived from the eclipses we adopt $\sigma_{\phi}=0.02$. For the radial velocity (RV) curves of the emission lines in CVs we adopt $\sigma_{\phi}=0.15$. The latter estimate reflect both random errors of measurements of broad emission lines of varying profile and possible systematic RV deviations to be discussed later. The errors consistent with Eq. (\ref{ea.1}) are coded (E) in Table \ref{t3}. We caution here that the above values are rough guesses to be used only as last resorts.
\subsubsection{Correletion or Red Noise Effects}
Direct least squares (LSQ) fit of CVs light curves result in claims of period errors substantially less than in Eq. (\ref{ea.1}). These are often artefacts resulting from improper treatment of intrinsic random flickering of a typical time scale $t_f$ of several minutes. In the time and frequency domains the flickering manifests respectively as correlation of consecutive observations and red noise. Thus any measurements repeated on a time scale $t_m$ short compared to $t_f$ yield no additional information on period (Schwarzenberg-Czerny, 1991). After correction for the correlation/red noise effects the LSQ errors increase by $\sqrt{t_f/t_m}$ and they became comparable to those of Eq. (\ref{ea.1}). Similar problems may arise if two different modulations in the light curve are difficult to separate causing mutual interference in the period estimation. The only difference is that now $P/2$ plays role of $t_f$.
We quote LSQ errors only when roughly consistent with (E). They are coded with (L).
\subsection{Systematic errors}
\subsubsection{Period changes}
In CVs systematic period modulations result from evolution of accretion disc and/or from solar cycle of the secondary star. Sparse observations of these effects often yield apparently semi-random scatter of the derived periods. To prevent over-interpretation we must caution here that neither the underlying physics nor telescope time allocation are random processes so that strictly speaking the resulting period estimates are not random variables. This said, we handled them as if they were random. Thus our prefered procedure was to derive periods by averaging the results from different observing runs and quoting half of their spread interval as their error. Occasionally period derivative was used to estimate the spread interval.
The errors derived from the spread interval are coded (S).
\subsubsection{Orbital period changes}
Studies of eclipses revealed that orbital periods of CVs fluctuate over time scales from several years
to several decades. Possible explanations involve solar-type cycle of the secondary star (Warner, 1995). However, these period changes $\delta_{\phi}<0.02$ for reasonable coverage should not affect orbital cycle count and they may be neglected in our analysis. In spectroscopic observations large random errors combine with large systematic orbital phase shifts, e.g. manifested by RV and eclipse phase differences.
The shifts reach $\delta_{\phi}<0.15$. They are likely caused by non-axsymmetric disk sources of the emission lines and they seem to depend on the luminosity/accretion status of a CV binary (Thorstensen et al. 1985, hereafter TSH). On one hand such shifts may prevent reliable orbital cycle count over gaps factor 3 larger (GF3L) than the base time interval. On the other hand consistent appearance of alias envelopes in the periodograms from different data sets suggests that likely cycle count errors do not exceed several over the whole data interval. Such errors are of negligible astrophysical consequence in the present context. 

Our point is well illustrated for TT Ari, a bright and well observed novalike CV. TSH derived its spectroscopic ephemeris by carefully attempting to elliminate any systematic effects
and using datasets of good quality spanning over 6 years. However, their stated period error was
consistent with $\sigma_{\phi}$ as small as $0.016$. After 21 years Wu et al. (2002) obtained another good sample of RV observations. They found TSH ephemeris shifted in phase by 0.35,
indicating period error underestimated and insecure cycle count. The updated period shift
is consistent with $\sigma_{\phi}\geq 0.09$ for TSH data. The new ephemeris by Wu et el. (2002) also does suffer from GF3L hence a $\pm1$ cycle count error over 21 years may not be entirely excluded. We stress again that such problems arise for a well observed star and carefull analysis. The true culprit are systematic effect in the accretion disc light sources. Insecure cycle count due to GF3L determinations are coded with (C).
\subsubsection{Period changes of disc humps} Negative and/or positive humps display periods close to the orbital one. They are observed as permanent features in some novalike stars or occasionally in dwarf novae, mainly during superoutbursts. They periods are known to vary in time. The variations may be systematic during superoutbursts or they manifest as period fluctuations between different superoutbursts and in permanent humps. The humps are thought to arise solely in an accretion disc. Their clocks are tied to the dynamical effects while any period changes reflect varying residual pressure effects. For sufficent data we estimated periods using (S) procedure and otherwise we had to rely on (E) values.
\section{Prewhitening with amplitude modulation}\label{a2}
While our prewhitening technique is rather simple and efficient, it seems to be rarely used hence we devote space to its brief description. We assume that the exact value of frequency $\omega$ is known, e.g. from a constant amplitude sine fit. We adopt time zero so that it corresponds the to maximum of the fitted sinusoid. Let $f(t)=\cos{\omega t}$ denotes a fixed cosine function and $P=2\pi/\omega$ is its period. For each data point $x_i$ obtained at time $t_i$ we select a running window extending over $t_i\pm P/2$. Within that window the data are fitted   with the function $a+bf(t)$ by least squares adjustement of the shift and scale parameters, $a$ and $b$ respectively. For the value of the modulated oscillation at time $t_i$ we adopt $y_i=a+bf(t_i)$.
For the next moment of time $t_{i+1}$ we select a new window and obtain new values of $a$ and $b$ and so on.

The above procedure may be efficiently implemented by noting that
\begin{equation}
  y_i=\frac{\left(n[xf]-[f][x]\right)f(t_i)+[x][f^2]-[xf][f]}{n[f^2]-[f]^2}
\end{equation}
where $[\cdot]$ denote sums of the corresponding values over the running window and $n$ is number of points in the window. These sums are easily adjusted after moving of the window by adding contributions from the new points and subtracting those from the omitted ones. Our particular selection of the window width of $P$ or its integer multiple is optimal for roughly uniform distribution of observations. Namely, it may be demonstrated that then the parameters $a$ and $b$ are uncorrelated, hence they suffer less from statistical errors.

\end{document}